\documentclass[printer]{tPHM2e}

\usepackage{rotating}

\begin{document}

\newcommand{\spd}{$sp$--$d$ }
\newcommand{\ef}{E_{\rm F}}
\newcommand{\du}{{\rm d}}
\newcommand{\e}{{\rm e}}
\newcommand{\Ang}{{\rm \AA}}

\newcommand{\be}{\begin{equation}}
\newcommand{\ee}{\end{equation}}
\newcommand{\ben}{\begin{eqnarray}}
\newcommand{\een}{\end{eqnarray}}
\newcommand{\beq}{\begin{equation}}
\newcommand{\eeq}{\end{equation}}
\newcommand{\B}{\mathrm{B}}
\newcommand{\NB}{\mathrm{NB}}
\newcommand{\RRe}{\mathrm{Re}}
\newcommand{\IIm}{\mathrm{Im}}

\doi{10.1080/1478643YYxxxxxxxx}

\markboth{G. Trambly de Laissardi\`ere, J. P. Julien, D. Mayou}
{Electronic transport in AlMn(Si) and AlCuFe quasicrystals: Break-down of the semiclassical model}

\title{
Electronic transport in AlMn(Si) and AlCuFe quasicrystals:\\
Break-down of the semiclassical model
}

\author{Guy TRAMBLY DE LAISSARDI\`ERE,$^{\dagger}$\thanks{$^\ast$Corresponding author. Email: guy.trambly@u-cergy.fr}$^\ast$
Jean-Pierre JULIEN$^{\ddagger \S}$  and Didier MAYOU$^{\S}$\\\vspace{6pt}
$^{\dagger}$ Laboratoire de Physique Th\'eorique et Mod\'elisation,
CNRS/Universit\'e de Cergy-Pontoise,\\
2 av. A. Chauvin, 95302 Cergy-Pontoise, France
\\
$^{\ddagger}$ Theoretical Division and CNLS,
LANL, Los Alamos, NM 87545, USA \\
$^\S$ Institut N\'eel, CNRS and Universit\'e Joseph Fourier, B\^at D,
25 av. des Martyrs,\\
B.P. 166, 38042 Grenoble Cedex 9, France\\
\received{December 17, 2007}}

\maketitle

\begin{abstract}
The semi-classical Bloch-Boltzmann theory is at the heart of our understanding of conduction in solids, ranging from metals to semi-conductors. Physical systems that are beyond the range of applicability of this  theory  are thus of fundamental interest. It appears that in quasicrystals and related complex metallic alloys, a new type of break-down of this theory operates. This phenomenon is related to the specific propagation of electrons. We develop a theory of quantum transport that applies to a normal ballistic law but also to these specific diffusion laws. As we show phenomenological models based on this theory describe correctly the anomalous conductivity in quasicrystals. Ab-initio calculations performed on approximants confirm also the validity of this anomalous quantum diffusion scheme. This provides us with an ab-initio model of transport in approximants such as  $\alpha$-AlMnSi and  AlCuFe 1/1 cubic approximant.

\end{abstract}

\section{Introduction}

Since the early 1990's experimental investigations have indicated that the conduction properties of several stable quasicrystals (AlCuFe, AlPdMn) are at the opposite of those of good crystals 
\cite{Poon92,Berger94,Grenet00_Aussois}. 
Within a decade a series of new quasiperiodic phases and approximants were discovered and intensively studied. These investigations taught us that indeed  electrons' and  phonons' properties could be deeply affected 
by this new type of order. There is now strong evidence that these non standard properties result from a new type of break-down of the semi-classical Bloch-Boltzmann theory of conduction.

Since the discovery of  Shechtman et al. \cite{Shechtman84} our view of the role of quasiperiodic order has evolved. For  electronic or phonon properties of most known alloys  it appears that the medium range order, on one or a few nanometers, is the real length scale that determines properties. This observation has lead the scientific  community to adopt a larger point of view and consider quasicrystals as an example of a larger class. This new class of Complex Metallic Alloys contains  quasicrystals, approximants and alloys with large and complex unit cells with possibly hundreds of atoms in the unit cell. 

In this paper we shall concentrate on ``the way electrons propagate'' 
in actual quasicrystal or in a complex metallic alloys. 
The main objective is to show that the non standard  conduction properties of some quasicrystals and related complex metallic alloys result from purely quantum effects and cannot be interpreted through the semi-classical theory of transport. 

In the Bloch-Boltzmann model the charge carriers are viewed as classical particles with velocity $V$ and charge $e$.
Their propagation is ballistic between two scattering events, separated by a characteristic time $\tau$, and they are scattered by static defects and/or phonons. This semi-classical description is valid if the size $L_{\rm WP}$ of the wave-packet is smaller than the distance of traveling between two scattering events $V \tau$, i.e. $V \tau > L_{\rm WP}$. 
But in a small velocity regime (SVR) such that $V \tau < L_{\rm WP}$, the semi-classical Block-Bolzmann model breaks down \cite{PRL06}. The SVR differs from another well known regime were the Bloch-Boltzmann model fails i.e. the regime with quantum interferences (weak or strong localization in disordered systems). 

We develop a theory of quantum transport that applies to a normal ballistic 
law but also to these specific diffusion laws \cite{PRL06,Mayou07_revue}.
This new formalism is combined with ab-initio band structure calculations for several approximant phases 
($\alpha$-AlMnSi, 1/1 AlCuFeSi) that share similar conduction properties with AlCuFe and AlPdMn icosahedral quasicrystals. As a result, we show that SVR explains the unconventional transport properties in quasicrystals and related phases.

\section{Experimental evidence of quantum diffusion in quasicrystals and related phases}
\label{ExperienceElectTranspPropert}
\subsection{Low density of states}

Experimentally a low density of states (DOS) at the Fermi
energy $E_{\rm F}$ is usually measured in
quasicrystals and their crystalline approximants.
For instance,
a density of states at
$E_{\rm F}$ reduced by the order of $1/3$ from
its free electrons value is measured in
i-AlCuLi and R-AlLiCu approximant
\cite{Poon92}.
The presence of the pseudogap in these phases
is confirmed by
NMR experiments~\cite{Hippert92} and X-ray measurements~\cite{Belin93}.

For icosahedral phases containing transition metal (TM) elements,
specific heat measurement indicate a DOS at
$E_{\rm F}$ the order of
$1/3$ of the free electron value for
i-AlCuFe
and $1/10$ for i-AlCuRu
and  i-AlPdRe~\cite{Poon92,Berger94}.
From X-ray spectroscopy the pseudogap
in the DOS is confirmed for many icosahedral quasicrystals in
the systems:
AlMn (metastable), 
AlMnSi, 
AlCuFe, 
AlCuFeCr, 
AlPdMn, 
AlCuRu, 
AlPdRe 
(see Ref. \cite{Belin02} and Refs. therein).
The pseudogap has been also measured in many approximants of
quasicrystals. For instance
R-AlCuFe~\cite{Hippert92,Berger94},
1/1 AlCuFeSi~\cite{Quivy96}
$\alpha$-AlMnSi~\cite{Berger94},
1/1 AlCuRuSi~\cite{Mizutani01,Mizutani04},
1/1 AlReSi~\cite{Takeuchi04}
have a DOS at $E_{\rm F}$ reduced by a similar factor
as in i-AlCuTM and i-AlPdMn.

\subsection{Conductivity: close to metal-insulator transition}

The first quasiperiodic alloys AlMn were metastable and they contained many structural defects. 
As a consequence they had conduction properties similar to those of amorphous metals with resistivities in the range 100--500\,$\mu\Omega$cm \cite{Berger94}. In 1986 the first stable icosahedral phase was discovered in AlLiCu. This phase was still defective. 
Although  its resistivity was higher (800\,$\mu\Omega$cm) it was still comparable to that of amorphous metals. The real breakthrough came with the discovery of the stable AlCuFe icosahedral phase, having a high structural order. The resistivity of these  well ordered systems were very high, of the order of 10\,000 $\mu\Omega$cm,  which gave a considerable interest in their conduction properties.  Within a few years several important electronic characteristics of these phases  were experimentally demonstrated. The conductivity presented a set of characteristics that were either  that of semi-conductors or that of normal metals. 

The density of states in AlCuFe is smaller than in Al, about one third of that of pure Al, but still largely metallic. 
Quasicrystals of high structural quality reveal unusual transport
properties \cite{Klein91,Poon92,Berger94,Grenet00_Aussois}.
For instance, one of the main features is the low conductivity
$\sigma_{4K}=100-200$~$\rm \Omega cm^{-1}$
for icosahedral AlPdMn and AlCuFe.
In particular  weak-localization effects were observed that are typical of amorphous metals. 
Yet the conductivity was increasing with the number of defects just as in semi-conductors.  

Another remarkable experimental result is 
the linear energy dependence of the optical conductivity 
of AlCuFe and the absence of  Drude peak \cite{Homes91,Burkov94_JPCM}.

In 1993 another breakthrough was the discovery of  AlPdRe  which had resistivities in the range of  $10^6\,\mu\Omega$cm  \cite{Pierce93_science,Berger93,Akiyama93,Delahaye99,Delahaye01,Delahaye03_JPCM,Rosenbaum04},
although the DOS still has a metallic character. 
This material displays a strong decrease of the conductivity when the temperature is reduced 
and the conductivity value can fall below 1\,$\rm (\Omega cm)^{-1}$ at 4\,K.
Although the behavior depends strongly on the composition and the preparation of the sample,
many authors \cite{Pierce93_science,Berger93,Akiyama93,Delahaye99,Delahaye01,Delahaye03_JPCM,Rosenbaum04} 
reported that AlPdRe quasicrystal are very close 
to the metal-insulator transition.
Three successive regimes are revealed \cite{Delahaye03_JPCM} as the temperature is increased 
to room temperature: a low temperature variable range hopping-like behavior, 
followed by a
Thouless regime and a high temperature critical regime.

It was also  experimentally shown that transition metal (TM) elements play an 
important role on the unusual transport properties of quasicrystals and 
related phases 
\cite{BelinMayou93,Berger93c,Mayou93b,Dankhazi93,GuyPRB95_2}.

Experimental measurements show that approximant phases
like $\alpha$-AlMnSi~\cite{Berger94},
1/1 AlCuFeSi~\cite{Quivy96},
R-AlCuFe~\cite{Berger94},
1/1 AlReSi~\cite{Tamura01,Takeuchi04}
etc.,
have transport properties similar to
those of quasicrystals AlPdMn and AlCuFe.
This suggests that the local atomic order
on the length scale of the unit cell,
{\it i.e.} $\rm 10-30~\AA$, determines their transport properties.
As atomic medium-range order of quasicrystals and
approximants are similar, it should also be important
in the understanding of transport properties of
quasicrystals.
This remark is confirmed by the fact that AlTM crystals
with a small unit cell
(typically less than 50 atoms in a unit cell)
do not exhibit such unusual transport properties.
In the following, crystals with small unit cell that do not exhibit transport 
properties similar to quasicrystals are called ``simple crystals''.

Finally it should be noted that in quasicrystals and approximants, the electron/phonons coupling is 
small and polarons \cite{Pascal} are not expected.

\subsection{Inverse Mathiessen rule}
\label{SecMathiessenRule}

The resistivity, $\rho=1/\sigma$, for crystals with
a small unit cell, increases  temperature $T$. 
Generally it follows  the Mathiessen rule:
\begin{equation}
\rho(T)=\rho_0 + \Delta{\rho}(T).
\end{equation}  
On the opposit, the resistivity of some quasicrystals and approximants
(AlPdMn, AlCuFe)
decreases when temperature increases , 
and their conductivity follows approximatively 
the so-called \textit{``inverse Mathiessen rule''} \cite{Mayou93,Berger94}:
\begin{equation}
\sigma(T)=\sigma_0 + \Delta{\sigma}(T).
\end{equation} 

Besides, after annealing sample, with a strong reduction of
the structural defects, the resistivity of quasicrystals and
approximants increases.
The relation between the particular transport 
properties of these phases
and their structure is still debated.
For AlPdMn quasicrystals,
J.J. Pr\'ejean and F. Hippert~\cite{Prejean02,Prejean06}
found that local defects might be related
with the occurrence of Mn atoms with localized magnetic moment
(see also F. Hippert and J.J .Pr\'ejean in this issue).
Thus, magnetic properties, transport properties and structural quality
are intimately linked for those complex phases.

\section{Main characteristics of electronic structure}

\subsection{Calculated density of states}

Electronic structure determinations have been
performed in the frame-work of 
density functional theory (DFT)
within the local density approximation (LDA)  by using the
self-consistent Tight-Binding (TB)
Linear Muffin Tin Orbital (LMTO) method
in the Atomic Sphere Approximation (ASA) \cite{Andersen75}.

The LMTO DOS of an $\alpha$-AlMn idealized approximant
has been first calculated by T. Fujiwara \cite{Fujiwara89,Fujiwara93}.
This original work shows the presence of a Hume-Rothery pseudogap
near the Fermi energy $\ef$ in agreement with experimental results
\cite{Poon92,Berger94}.
Other approximants such as 1/1-AlCuFe(Si) exhibit also a pseudogap near $\ef$
(see Refs. \cite{Mizutani01,PMS05} and Refs. in there)

The role of the transition metal (TM, TM = Ti, V, Cr, Mn, Fe, Co, Ni) element in the pseudogap formation
has  been shown from  ab-initio calculations
\cite{GuyPRB95,PMS05}.
Indeed the formation of the pseudogap  results from
a strong sp--d coupling associated to an ordered sub-lattice
of TM atoms.
Just as for Hume-Rothery phases, a description of the band energy
can be made in terms of pair interactions.
It was shown that
a medium-range TM--TM interaction mediated by
sp(Al)--d(TM) hybridization plays a determinant role
in the occurrence of the pseudogap 
\cite{Pasturel86,Friedel87,Fujiwara89,Fujiwara93,ZouPRL93,GuyEuro93,GuyPRB95,PMS05,Guy03,GuyPRL00,Guy04_ICQ8,Virginie2,Hippert99}.
It is thus essential to take into account the chemical nature of
elements to analyze the electronic properties of approximants. 
The electronic structures of simpler crystals such as
 $\rm Al_6Mn$, $\rm \omega - Al_7Cu_2Fe$, $\rm Al_{13}Fe_4$,
 $\rm Al_{12}Mn$, present \cite{PMS05} also a pseudogap near $\ef$
which is less pronounced than in complex approximant phases. 

\subsection{Electron localisation by atomic clusters}
\label{Sec_cluster}
As for the local atomic order, 
one of the characteristics of the quasicrystals and approximants, 
is the occurrence of atomic clusters on a scale of 10--30 $\rm \AA$ \cite{Gratias00}.  
Nevertheless the clusters are not well defined because some of them overlap,  
and in addition there are a lot of so-called glue atoms. 
The role of clusters has been much debated in particular by 
C. Janot \cite{Janot94}
and in Ref. \cite{GuyPRB97}. 
It is realistic to consider a model of  clusters that are not isolated 
but  are embedded in metallic medium.   

As shown in Refs.\cite{GuyPRB97,GuyICQ6}, the variation
$\Delta n_{\rm cluster}(E)$ of the DOS due to a TM cluster exhibits strong deviations from the 
Virtual Bound States (1 TM atom in metallic medium).
Indeed several peaks and shoulders appear. The width $\delta E$ of the most narrow peaks 
($\delta E \simeq 10 - 100$\,meV)
are comparable to the fine peaks of the calculated DOS in the approximants. 
Each peak indicates a resonance due to the scattering by the cluster. 
These peaks correspond to states ``localized''  
by the cluster. 
They are not eigenstate, they have finite lifetime of the order of $\hbar / \delta E$, 
where $\delta E$ is the width of the peak. 
Therefore, the stronger the effect of the localization by cluster is, the narrower is the peak. 
A large lifetime is the proof of a localization, 
but in real space these states have a quite large extension on length scale of the cluster 
($\sim 50$ atoms).


This effect is a multiple scattering effect, 
and it is not due to an overlap between d-orbitals because TM atoms are not first neighbors 
(TM -- TM first neighbors distance is $\sim 4.8\,\Ang$).
We have also shown that these resonances are very sensitive to the geometry of the TM cluster.
For instance, they disappear quickly when the radius of the TM icosahedron increases, 
and they are strongly reduce by vacancy. Therefore transport properties should be very sensitive 
to the atomic positions of TM atoms and to the chemical composition.

\section{Calculated transport properties}

\subsection{Atomic structure model for approximants}

To illustrate the quantum diffusion in approximants of quasicrystals we consider two phases:
the $\alpha$-AlMnSi approximant and a model for AlCuFeSi 1/1 cubic approximant.

For the  $\alpha$-AlMnSi phase, 
we use the experimental atomic structure
\cite{Sugiyama98}
with the Si positions proposed by E. S. Zijlstra and S. K. Bose
\cite{Zijlstra03} for the composition
$\alpha$-Al$_{69.6}$Si$_{13.0}$Mn$_{17.4}$.
This phase contains 138 atoms in a cubic unit cell:
96 Al atoms, 18 Si atoms, and 24 Mn atoms.

V. Simonet {\it et al.}~\cite{Simonet05_AlCuFe}
refined experimentally the atomic structure and the chemical
decoration of Al--Cu--Fe--Si 1/1 cubic approximants.
The authors give a revised description of the structure of  
$\alpha'$-$\rm Al_{71.7}Si_7Cu_{3.8}Fe_{17.5}$ phases
and  $\alpha$-$\rm Al_{55}Si_7Cu_{22.5}Fe_{12.5}$ phase.
$\alpha'$-phase has a chemical
decoration similar to that 
of $\alpha$-Al--Mn--Si, 
whereas the structure and the composition of the $\alpha$-phase is 
different.
It is characterized by several Wyckoff sites with mixed occupancy between
Al/Cu, Al/Fe and Cu/Fe.
As an example, we used
this  structure to calculate the LMTO DOS for
phase with the composition
$\rm Al_{78}Cu_{48}Fe_{13}$ in a cubic unit cell.

\begin{figure}[]
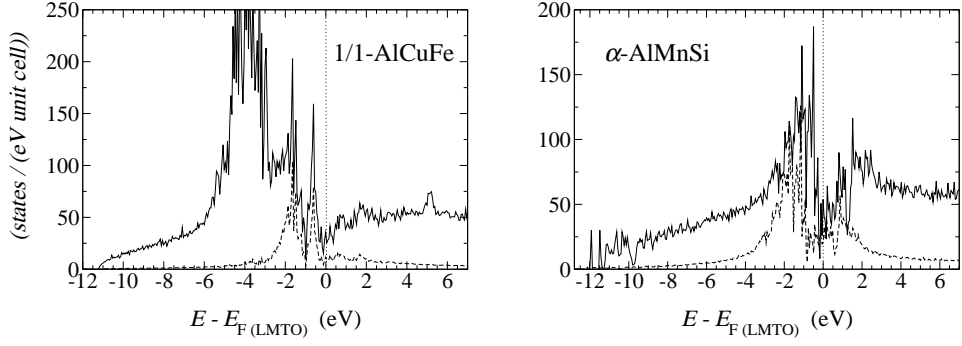

\begin{center}
\includegraphics[height=4.5cm]{DOS_aAlCuFe_dos10.eps}
~~~~~~\includegraphics[height=4.5cm]{sugi_dAlSi_Zijl_b_6_all-Mn.eps}
\caption{Ab-initio LMTO DOS in $\alpha$-Al$_{69.6}$Si$_{13.0}$Mn$_{17.4}$
 and 
$\rm Al_{78}Cu_{48}Fe_{13}$ 1/1-cubic approximant: 
(line) total DOS and 
(dashed line) local $TM$ DOS ($TM =$ Mn or Fe).
\label{LMTO_alpha_AlCuFe}}
\end{center}
\end{figure}

In Fig. \ref{LMTO_alpha_AlCuFe}, the total DOS $n(E)$ of these phases 
are presented. A pseudogap near $E_{\rm F}$ is clearly seen.
Following the Hume-Rothery condition, 
it is expected that the most realistic value of
$E_{\mathrm{F}}$  corresponds to
the minimum of the pseudogap.
As shown previously for AlCuFe model approximant~\cite{GuyPRB94_AlCuFe},
the positions of Fe atoms have strong effects on the DOS near the
$E_{\rm F}$, and thus on the pseudogap and the stability.
A detailed analysis \cite{Zijlstra04}
of a modified Cockayne model after a structural
relaxation confirms the effect of the TM positions on the
stability.
Results presented here for 1/1~AlCuFe give 
thus a good qualitative example of the quantum diffusion in approximants, but a more detailed studies
of the composition effect are still necessary 
to obtain quantitative results in AlCuFe(Si) approximants.

\subsection{Ab-initio calculations of the quantum diffusion}

We now present calculations of  quantum diffusion in perfect crystalline systems. 
In literature, several calculations have already been done from ab-initio studies 
(see for instance Refs. \cite{Roche96,Fujiwara96,Roche98,Krajci02}). They give indication 
of non-ballistic diffusion \cite{Roche96,Fujiwara96,Roche98,Triozon02,Krajci02,Bellissard03}.
In our approach of quantum diffusion, 
the main quantities are the {\it velocity correlation function}:
\ben
C(E,t) = \Big\langle V_x(t)V_x(0) + V_x(0)V_x(t) \Big\rangle_E
= 2\,{\rm Re}\, \Big\langle V_x(t)V_x(0) \Big\rangle_E ,
\label{EqAutocorVit}
\een
where $V_x$ is the velocity operator,
and the {\it square spreading} $\Delta X^2(E,t)$ 
of electronic states with energy $E$ at time $t$ \cite{MayouPRL00}.
These two quantities are simply related by the relation:
\ben
\frac{\du}{ {\du} t} \Delta X^2(E,t) = \int_0^{t}C(E,t'){\du} t'.
\label{RelXC}
\een
In crystals, these quantities can be decomposed 
in a ballistic contribution (Boltzmann term) and a 
non-ballistic contributions (non-Bolzmann term):
\begin{equation}
C(E,t)= 2~ V_{\mathrm{B}}(E)^2  + C_{\NB}(E,t) 
~~~{\rm and}~~~
\Delta X^2(E,t) = V_{\mathrm{B}}(E)^2 t^2 + \Delta X_{\mathrm{NB}}^2(E,t),
\label{Eq_DeltaX2}
\end{equation}
where $V_\mathrm{B}$ is 
the Boltzmann velocity at energy $E$.
The ballistic terms:
\be 
C_{\mathrm{B}} =2 V_{\mathrm{B}}(E)^2 
{\rm ~~and~~}
\Delta X_{\mathrm{B}} = V_{\mathrm{B}}(E)^2 t^2
{\rm ~~with~~}
V_{\mathrm{B}}(E)^2 =  \langle 
|\langle n\vec k | V_x | n\vec k \rangle |^2 \rangle_{E_n=E},
\label{Eq_termeBoltzman}
\ee
are due to intraband contributions;
and the non-ballistic terms $C_{\NB}(E,t)$, $\Delta X_{\mathrm{NB}}^2(E,t)$ 
are due to the interband contributions:
\begin{equation}
\Delta X^2_{\mathrm{NB}}(E_{\mathrm{F}},t) = 2 \hbar^2
~\Bigg\langle
\sum_{m \,(m\neq n)}
\frac{1 - \cos\Big((E_n-E_m)\frac{t}{\hbar} \Big)}{(E_n-E_m)^2} 
\Big| \langle n\vec k | V_x | m\vec k \rangle \Big|^2
\Bigg\rangle_{E_n=E_{\mathrm{F}}}.
\label{Calcul_DeltaX2}
\end{equation}
In equations (\ref{Eq_termeBoltzman}) and (\ref{Calcul_DeltaX2}),
$| n\vec k \rangle$ is an eigenstate with energy $E_n$.
$\Delta X_{\rm NB}^2(E)$ is the average spreading of the state within a unit cell.
Thus a relation exists between $\Delta X_{\rm NB}^2(E)$ 
and the length $L_c$ of the unit cell in the chosen direction namely \cite{Mayou07_revue}
\ben
\Delta X_{\rm NB}^2(E,t) \leq
\left(\frac{L_c}{2}\right)^2.
\label{bound}
\een

From self-consistent LMTO eigenstates,
we compute the velocity correlation function  
$C(E,t)$ and $\Delta X(E,t)$ for crystals (approximant and simple crystals).
In equations ($\ref{EqAutocorVit}$) ($\ref{Calcul_DeltaX2}$), 
the average $\big\langle ~ \big\rangle_{E}$
on states with the same energy $E$
is obtained by taking the eigenstates for each
$\vec{k}$ vector with an energy $E_n(\vec{k})$ such as
$E-\Delta E /2<E_n(\vec{k})<E+\Delta E/2$.
$\Delta E$ is the energy resolution of the calculation. The calculated
$C(E,t)$ is sensitive to the  number $N_k$ of $\vec{k}$
vectors in the first Brillouin zone
when $N_k$ is too small. 
Therefore $N_k$ is increased
until $C(E,t)$ does not depend significantly on $N_k$.
For 1/1-AlCuFe and $\alpha$-AlMnSi,
$\Delta E = 0.0272$\,eV 
and $N_k = 32^3$.

\subsection{Results}

\begin{figure}[]
\begin{flushright}

\includegraphics[width=8cm]{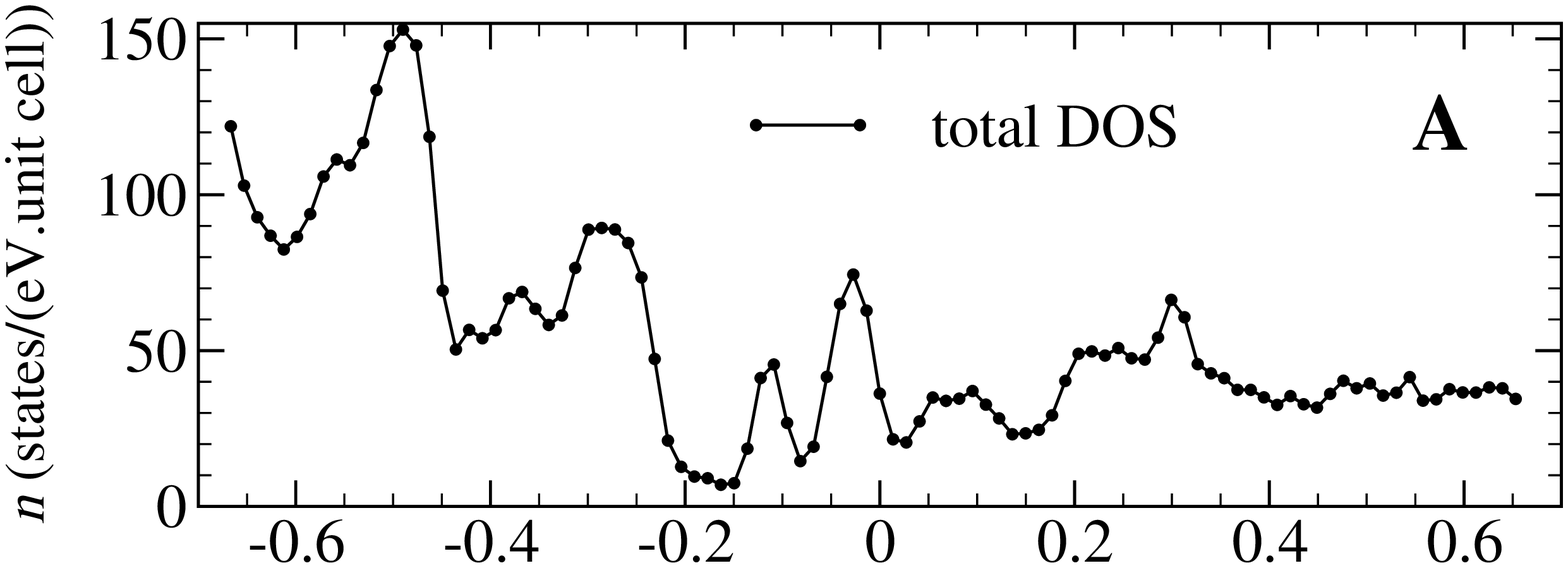}\hspace{5cm}

\vspace{.2cm}
\includegraphics[width=8.3cm]{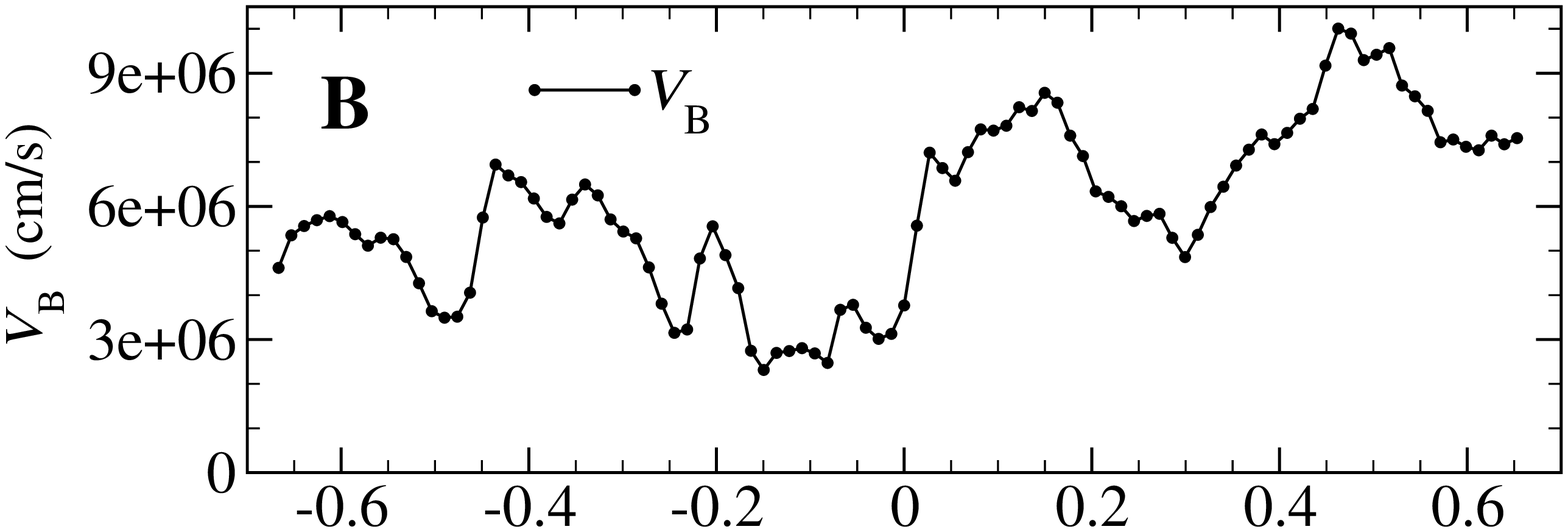}\hspace{5cm}

\vspace{.2cm}
\includegraphics[width=8.1cm]{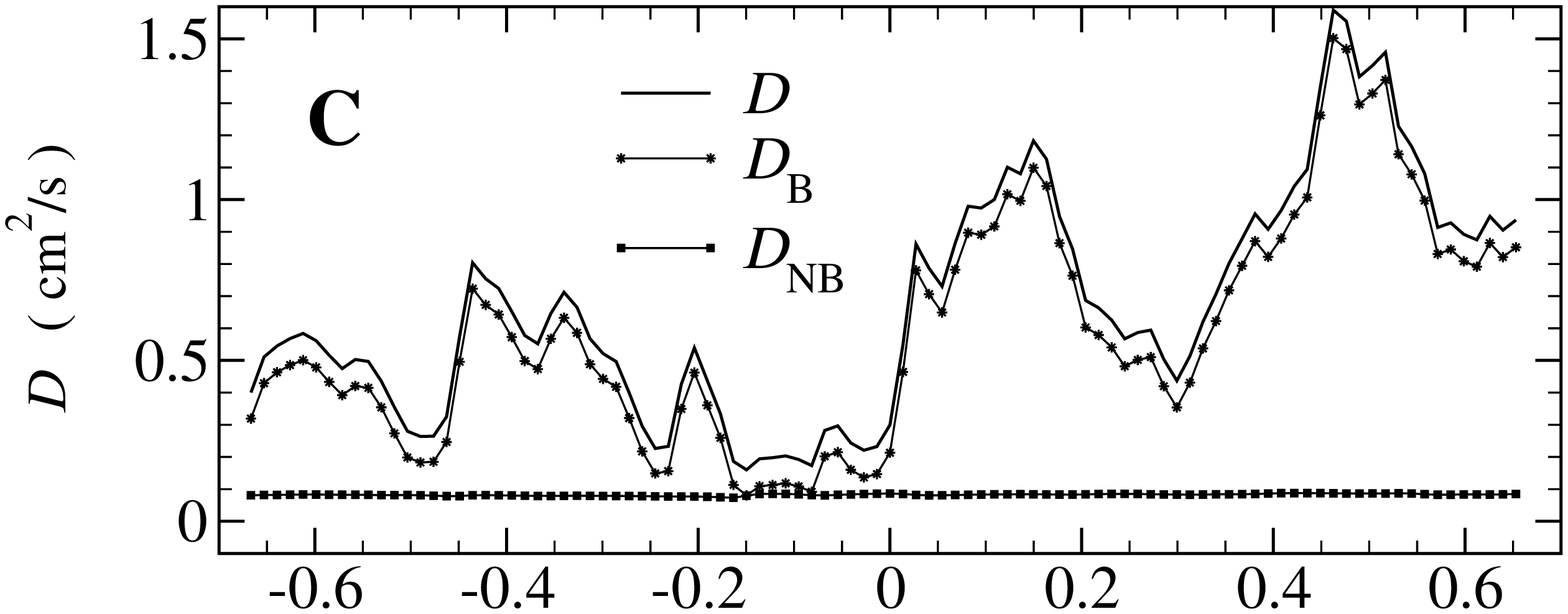}\hspace{5cm}

\vspace{.2cm}
\includegraphics[width=8cm]{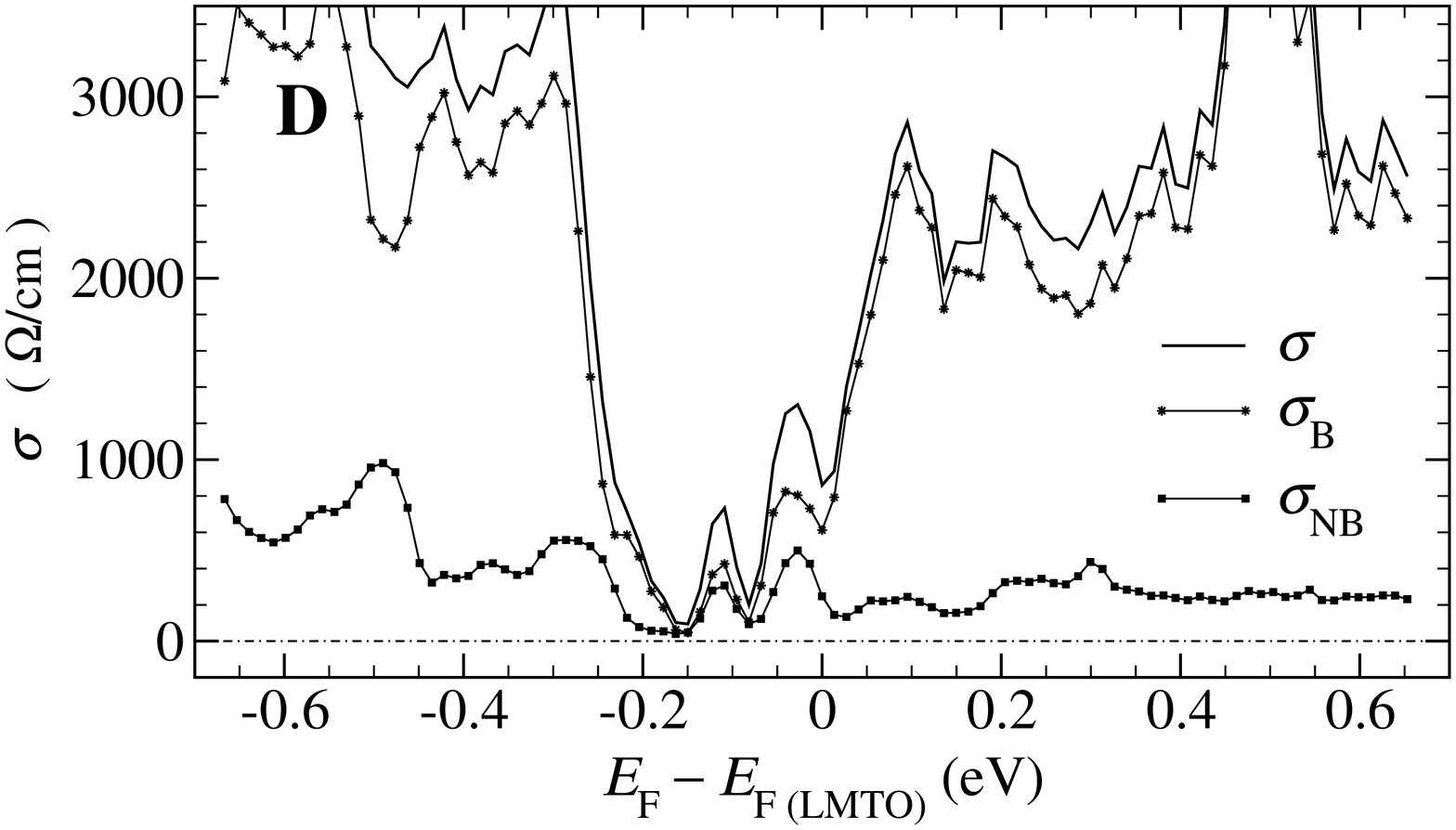}\hspace{5cm}

\end{flushright}

\vspace{.2cm}
\caption{ \label{Fig_DOS_VB_Dif_Sig}
({\bf A}) LMTO total DOS $n$, 
({\bf B}) Bolztmann velocity $V_{\B}$, 
({\bf C}) diffusivity $D = D_{\B}+D_{\NB}$, 
and
({\bf D}) conductivity $\sigma = \sigma_{\B} + \sigma_{\NB}$, 
in the
cubic approximant $\alpha$-Al$_{69.6}$Si$_{13.0}$Mn$_{17.4}$.
Points are the calculated values and lines are guides for the eyes.
$D$ and $\sigma$ are calculated for $\tau = 1.5 \times 10^{-14}$\,s.
}
\end{figure}

The Boltzmann velocity (intra-band velocity) $V_{\B}$
in $\alpha$-AlSiMn versus the Fermi energy $\ef$,
is shown on figure  \ref{Fig_DOS_VB_Dif_Sig}.B.
Similar results are obtains in 1/1\,AlCuFe.
These values for approximants are
also similar to the original work of  T. Fujiwara et al.
\cite{Fujiwara93,GuyPRB94_AlCuFe,GuyPRBAlCuCo}.
$V_{\B}$ in approximants varies very rapidly with a small variation of $\ef$, which shows 
the crucial effect of the chemical composition on transport properties.
The minimum value of $V_{\B}$ is about $\rm 2.7 \times 10^6\,cm.s^{-1}$, 
it corresponds to minimum in the DOS $n(E)$ (figure \ref{Fig_DOS_VB_Dif_Sig}.A).
In simple
crystals Al (f.c.c.) and  cubic $\rm Al_{12}Mn$:
$V_{\rm }= 9 \times 10^{7}$ and $4 \times 10^{7}\,cm.s^{-1}$, 
respectively \cite{ICQ9}.
The reduction of
$V_{\rm F}$ in the approximant phases with respect to simple
crystal phases 
shows the importance of a quasiperiodic
medium-range order (up to distances equal to 12--20\,$\rm \AA$). 
This leads to a very small Boltzmann conductivity in approximants.

\begin{figure}[]
\begin{center}
\includegraphics[width=8cm]{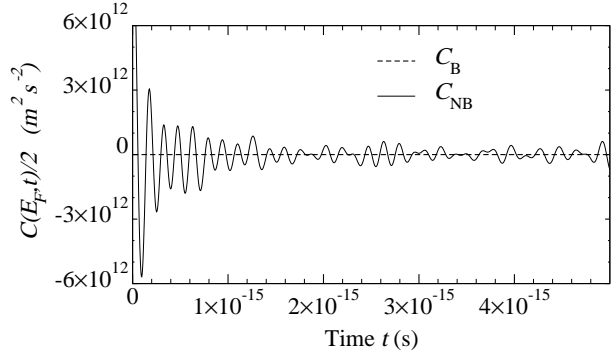}
\caption{Velocity correlation function $C(\ef,t)$ 
in  $\alpha$-Al$_{69.6}$Si$_{13.0}$Mn$_{17.4}$
versus large time $t$.
Dashed lines are the corresponding Boltzmann velocity
correlation function $C_{\rm B}(\ef,t)=2v_{\rm F}^2$.
\label{Fig_vx2}}
\end{center}
\end{figure}

The velocity correlation function
$C(\ef,t) = C_{\B} +C_{\NB}$ for the $\alpha$-AlSiMn
is shown in
figure \ref{Fig_vx2}. 
In the case of Al and other simple crystal, $C(\ef,t)$ is almost always positive,
and the Boltzmann value is reached
rapidly when $t$ increases \cite{ICQ9}.
But for many $t$ values the velocity correlation functions $C(\ef,t)$ of approximants are
negative.
This means that at these times the phenomenon of
{\it backscattering} occurs \cite{ICQ9,BouquinJP05,Mayou07_revue}.
The transport properties
depend on the average value of $C(\ef,t)$ on a time scale equals to the
scattering time $\tau$ \cite{MayouPRL00,Triozon04}.
Therefore, in simple crystals, the backscattering
(negative value of $C(\ef,t)$) should have a negligible effect on
the transport properties, whereas this effect must be determinant
for approximants.

\begin{figure}[]
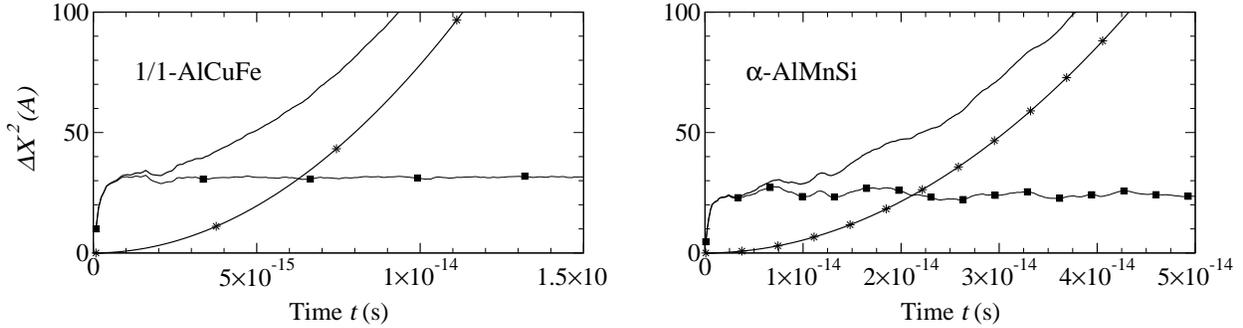

\begin{center}
\includegraphics[height=4.3cm]{X2_aAlCuFe_e-.01D.001.eps}
~~\includegraphics[height=4.3cm]{X2_alpha_Z18b_e-.011.eps}
\end{center}
\caption{ \label{Fig_X2}
Square spreading 
$\Delta X^2$
of electrons states with Fermi energy 
$E_{\mathrm{F}}$ versus time $t$, 
in 1/1 AlCuFe phases 
and $\alpha$-Al$_{69.6}$Si$_{13.0}$Mn$_{17.4}$:
(simple line) total $\Delta X^2$, (line with stars) Bolzmann term $\Delta X_{\mathrm{B}}^2$
and (line with square) non-Bolzmann term $\Delta X_{\mathrm{NB}}^2$.
}
\end{figure}
The phenomenon of
backscattering is associated to unusual quantum diffusion.
It is
illustrated on the plot of the average spreading of states
$\Delta X^2$ versus time $t$ (figure \ref{Fig_X2}).
The non-Boltzmann contribution, $\Delta X^2_{\rm NB}$,
increases very rapidly and saturates to a maximum value
of the order of the square size of the unit cell.
In approximants, at small time $t$,  
$\Delta X^2_{\rm B}$ is smaller than in simple phases due to a very small velocity
$V_{\rm B}$.

Thus approximants are a non-conventional metal at these
time scale i.e. when the scattering time is $\tau < \tau^*$
where $\tau^*$, the limit of two regimes (see next section), is around
$1.5\times 10^{-14}\,$s and $6 \times 10^{-15}\,$s, 
in $\alpha$-AlMnSi and 1/1~AlCuFe,  respectively.
At realistic scattering times scale for approximants, typically  $\sim 10^{-14}$ \cite{Mayou93},
both terms $\Delta X^2_{\B}$ and $\Delta X^2_{\NB}$ have 
the same magnitude ($\tau \lesssim \tau^*$);
whereas
in normal crystals, the $\Delta X^2_{\NB}(t)$ 
term is negligible with respect
to the Boltzmann term $\Delta X^2_{\B}(t)$ because $\tau \gg \tau^*$.

\section{Static conductivity of approximants in relaxation time approximation}

For finite  temperature, the effect of the   
Fermi-Dirac distribution on transport properties
was studied in the 
literature \cite{Macia02-04,Solbrig00,Haussler02,Macia07}. 
But, theses analyzes could not
explain the unconventional conduction  of quasicrystals and 
related alloys 
(very high resistivity at low temperature, 
and conductivity that increases strongly 
when 
defects or temperature increases).
In the following,
the Fermi-Dirac distribution function is taken equal to its zero temperature value.
This is valid provided that the electronic properties vary
smoothly on the thermal energy scale  $k_{\B}T$. 
However,
the effect of defects and temperature
on the conductivity is taken into account via
the Relaxation Time Approximation (RTA).
A scattering time $\tau$ is defined as the average 
time between two collisions of an electron 
with static impurities and/or phonons.
$\tau$ includes both elastic and inelastic scatterings, it 
decreases when temperature or static defects increase.

Within the RTA, the velocity correction function $C'(E,t)$ 
with disorder are related to $C(E,t)$ without disorder through \cite{MayouPRL00}:
\ben
C'(E,t) = C(E,t) \e^{-|t| / \tau},
\label{Relaxationp2}
\een
and
the dc-diffusivity at energy $E$ is given by 
 \ben
D(E) = \frac{1}{2} \, 
\int_0^{+\infty} \e^{-t/\tau} C(E,t) {\du}t = D_{\B}(E)+D_{\NB}(E),
\label{D(C0)p}
\een
where Boltzmann diffusivity is $D_{\B}(E) = V_B^2(E) \tau$, 
and the non-Boltzmann term is
\ben
D_{\NB}(E) = \frac{1}{2} \,  
 \frac{1}{\tau^2}
\int_0^{+\infty} \e^{-t/\tau} \Delta X_{\NB}^2(E,t) {\du}t.
\label{DNB21}
\een
As  figure \ref{Fig_DOS_VB_Dif_Sig}.C shows,
$D_{\mathrm{NB}}$ is almost independent
on $E$, whereas the
$D_{\mathrm{B}}$ values depend strongly on $E$,
as $V_{\B}$ value depends on $E$.
\begin{figure}[]
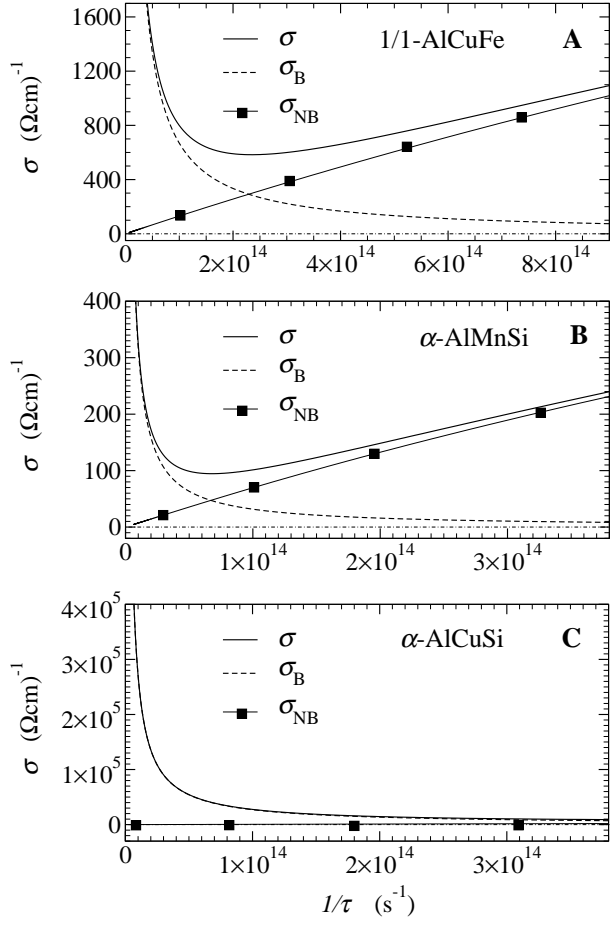

\begin{flushright}

\includegraphics[width=8cm]{aAlCuFe_d10_3t10_T_sig_e-.01D.001.eps}\hspace{4cm}

\vspace{.2cm}
\includegraphics[width=8cm]{sugiZ18bd10_3_T_sig_e-.011D.001.eps}\hspace{4cm}

\vspace{.2cm}
\includegraphics[width=8cm]{sugiAlCuSI_t10_T_sig_e-012D001.eps}\hspace{4cm}

\end{flushright}

\caption{ \label{Fig_Conduc_inv_tau}
Ab-initio dc-conductivity $\sigma$ ($\sigma = \sigma_{\B} + \sigma_{\NB}$) versus
inverse scattering time $1/\tau$, 
in 
({\bf A}) 1/1 approximant AlCuFe,
({\bf B})  approximant $\alpha$-Al$_{69.6}$Si$_{13.0}$Mn$_{17.4}$,
and
({\bf C}) hypothetical approximant $\alpha$-Al$_{69.6}$Si$_{13.0}$Cu$_{17.4}$.
}
\end{figure}
The dc-conductivity is 
\ben
\sigma(\ef)=\e^2n(\ef)D(\ef) =  \sigma_{\B}(\ef) + \sigma_{\NB}(\ef)
\een
where $e$ is the charge of electron.
The Boltzmann term is given by the Einstein relation, 
$\sigma_{\B}(\ef)=\e^2n(\ef) V_{\B}^2 \tau$. It is proportional 
to scattering time $\tau$ and it is small in approximants.
The non-Boltzmann term  $\sigma_{\NB}(\ef)$ is calculated from 
$\Delta X_{\NB}$ via equation (\ref{DNB21}).
As $X_{\NB}$ is almost constant for significant $t$ values 
(figure \ref{Fig_X2}), one obtains 
that  $\sigma_{\NB}(\ef)$ is almost proportional to $1/\tau$. 
Therefore~:
\be
\sigma ~=~ e^2 n V^2_{\B} \tau ~+~ e^2 n \frac{L^2(\tau)}{\tau} 
{\rm ~~with~~} L(\tau) \simeq L,
\label{Eq_sigma_B_NB}
\ee
where $L$ is bound by $L_c/(2 \sqrt{2})$ where $L_c$ is the size of the unit cell along which $\sigma$ 
is calculated (equation(\ref{bound})).
Roughly speaking, $L$ is the spreading of electron states in each cell. In simple
crystals, $L$ is small, but in approximants it is larger. As an electronic 
localisation by cluster exist in approximants (Sec. \ref{Sec_cluster}), 
$L$ should be close to 
the size of cluster. A large  value of $L$ is thus a consequence 
of a quasiperiodic local order.
The minimum value of $\sigma(\tau)$ is obtains for 
$\tau = \tau^* = L/V_B$. 

The predicted static conductivity (dc-conductivity) of the 
$\alpha$-AlMnSi and 1/1 AlCuFe approximants
assuming the value of the Fermi energy at the minimum of the pseudogap,
are shown  figure \ref{Fig_Conduc_inv_tau}
versus the inverse scattering time.
Two regimes appear clearly: 
\begin{itemize}
\item 
A metallic regime (Boltzmann regime), for $\tau >\tau^*$,
where $\sigma$ is almost proportional to $\tau$, and then
$\sigma$ decreases with disorder (static disorder or temperature)
as for simple crystals (Mathiessen rule).

\item 
An ``insulating like'' regime (non Boltzmann regime), 
for $\tau <\tau^*$,
where 
$\sigma$  is almost proportional to $1/\tau$,  
and then $\sigma$
increases with disorder
as observed experimentally for approximants and quasicrystals 
(inverse Mathiessen rule).
It should be noted that in this cases the system is always metallic (no gap in the DOS), but 
its conductivity is ``insulating like''.
\end{itemize}

For $\alpha$-AlSiMn,
realistic $\tau$ values
\cite{Mayou93} correspond to the  ``insulating like'' regime.
Therefore,
$\sigma$ increases when
defects or temperature increases.
$\sigma$ varies from 100~($\Omega$\,cm)$^{-1}$ for
$\tau = 1.5\times 10^{-14}$\,s,
to $\sim 2000$~($\Omega$\,cm)$^{-1}$ for
$\tau = 10^{-15}$\,s.
This is consistent with experimental results
in $\alpha$-AlMnSi:
$\sigma(4~K) \simeq 200$~($\Omega$\,cm)$^{-1}$
and $\sigma(900~K) \simeq 2000$~($\Omega$\,cm)$^{-1}$ 
and with standard estimates for the scattering time 
in these systems \cite{Berger94}.\\

Within the relaxation time approximation used here, 
the optical conductivity  $\sigma(\omega)$ can also be calculated as 
the sum of two terms \cite{PRL06,Mayou07_revue}.
The Boltzmann contribution
gives rise to the so-called Drude peak
and the non Boltzmann conductivity gives rise to a nearly frequency independent contribution. 
The absence of Drude peak in quasicrystals and approximant is thus explained by the insulating 
like regime for realistic $\tau$ values.\\

To evaluate the effect of TM elements
on the conductivity, we have considered an
hypothetical   $\alpha$-Al$_{69.6}$Si$_{13.0}$Cu$_{17.4}$
constructed by
putting Cu atoms in place of Mn atoms in the actual
$\alpha$-Al$_{69.6}$Si$_{13.0}$Mn$_{17.4}$ structure.
Cu atoms have almost
the same number of sp electrons as Mn atoms,
but their d DOS is very small at $\ef$.
Therefore in $\alpha$-Al$_{69.6}$Si$_{13.0}$Cu$_{17.4}$,
the effect of  sp(Al)-d(TM) hybridization on electronic states
with energy near $\ef$ is very small.
As a result, the pseudogap disappears in total DOS \cite{GuyPRB95,PMS05},
and the dc-conductivity is now metallic 
as shown on figure~\ref{Fig_Conduc_inv_tau}.C. \\

Equation (\ref{Eq_sigma_B_NB}) allows also to understand transport in quasicrystals (non periodic phases).
Indeed in quasicrystals $V_\B$ should be very small, 
and $L_c$ is equal to infinity. But a finite value of $L$ 
is possible depending on the electron localisations by the quasiperiodic structure.
Therefore in quasicrystals, the non Boltzmann term dominates and an ``insulating like''
regime is expected.

\section{Conclusion}
\label{SecConclusion}

In this paper we calculated quantum diffusion and  electronic conduction 
properties in two 1/1 approximants. 
We found deviations from the standard ballistic propagation in good 
agreement with experimental measurements.
The anomalous diffusion mode is related to a tendency to localization 
and to a phenomenon of backscattering which is well known in disordered systems. 
The phenomenon of backscattering is the fact that an impulse of electric field creates a 
current density which is opposite to the electric field at large time. 
Backscattering is associated  with an increase of conductivity with frequency and disorder. 
The physics of phonons in quasicrystals could also be affected by the anomalous diffusion phenomenon. 
In particular it has been argued that the heat conductivity could 
be sensitive to this effect \cite{Mayou00_Aussois}.

The concepts developed here open also a new  insight in the physics 
of correlated systems. Indeed recent studies of some heavy fermions 
or polaronic systems 
\cite{Vidhyadhiraja05,Fratini03,Fratini06,Cappelluti07},  
where charge carriers are also slow, show that their conduction properties 
present a deep analogy with those described here. 
In particular a transition from a metallic like regime 
at low temperature where scattering is weak  to an insulating 
like regime at higher temperature with a stronger scattering is observed.

\section*{Acknowledgements}

The works presented in the paper have been done
since the 90's.
Our work owes much to the discussions with
T. Fujiwara,
J. Bellissard,
J. Friedel,
N. W. Ashcroft.
We are very grateful to
many colleagues with whom we had collaborations during this time:
C. Berger,
F. Cyrot--Lackmann,
J. Delahaye,
T. Grenet,
F. Hippert,
T. Klein,
L. Magaud,
J. J. Pr\'ejean,
P. Qu\'emerais,
S. Roche,
F. Triozon.
The computations have been performed at the
Service Informatique Recherche (S.I.R.),
Universit\'e de Cergy-Pontoise.
Part of the numerical results has been
obtained by using  the Condor Project
(http:$/\!/$www.condorproject.org$/$).
GTL thanks Y. Costes, S.I.R.,
for computing assistance.
JPJ acknowledges the LANL group for whom hospitality
and DGA under contact 07.60.028.00.470.75.01.

%
%
%
%
\bibliographystyle{plain}

\end{document}